\documentclass[12pt]{article}
\usepackage{latexsym,amsfonts,amsmath,amsthm,amssymb}

\title{On Fundamental Principle of Quantum Theory}
\author{Andrei Khrennikov\footnote{
International Center for Mathematical
Modeling in Physics and Cognitive Sciences,Email:Andrei.Khrennikov@msi.vxu.se;
Suuported by the EU-network on Quantum Probability and its applications.}\\
MSI, University of V\"axj\"o, S-35195, Sweden}

\begin{document}
\maketitle

\begin{abstract}
We propose the principle, the law of statistical balance for basic physical observables,
which specifies quantum statistical theory  among all other
statistical theories of measurements. 
It seems that this principle might 
play in quantum theory the role 
that is similar to the role of Einstein's relativity principle.
\end{abstract}

\section{Introduction}
We start our paper with the citation from the 
work of A. Zeilinger [1]:

\medskip

`It so happened that almost all relativistic
equations which appear in Einstein's  publication of 1905 were known already before,
mainly through Lorentz, Fitzgerald and Poincare - simply as an
attempt to interpret experimental data quantitatively. But only Einstein 
created the conceptual foundations, from which, together with the constancy of the velocity of light,
the equations of the relativity arise. He did this by introducing the principle of relativity...'--
[2]. 

\medskip

A. Zeilinger was looking for a new quantum paradigm. He correctly 
underlines that the situation in quantum theory, especially large diversity of
interpretations, is not so natural, see also e.g. [3]. That in the opposite to e.g. theory of 
relativity, there is no quantum analogue of the {\it fundamental principle.} 
Following to A. Zeilinger [1], we are looking for an analogue of the fundamental
principle that would play the role of Einstein's relativity principle. 

In this paper I try to  propose  a candidate for such a ``quantum fundamental principle''. We start
with the remark that quantum formalism is the statistical theory; a theory that
deals with probabilities of events\footnote{However, as we have already discussed,
the existence of such a probabilistic description does not imply the impossibility 
realistic ontic description, [26].}.
Therefore any ``fundamental quantum principle''
(if it exists at all) should be a statement about probabilities. So we are looking
for a fundamental statistical principle that would induce quantum formalism.

Our statistical analysis of foundations of quantum theory is based on a so called {\it contextual
approach} to quantum probabilities, see [4], [5], cf. L. Ballentine [6]--[8], L. Gudder 
[9]--[11].\footnote{I also remark that my contextual investigations were initially stimulated by 
investigations of E. Prugovecki [12], [13] on the role of statistical and individual errors in quantum and classical
measurements.}

We also remark that ideas which are similar to Zeilinger's idea to find a `quantum fundamental principle'
(or principles)
are intensively discussed by many authors. In particular, extended discussions took
place at the V\"axj\"o-Conference on foundations of quantum theory, see [14].
I would like to mention papers of C. Fuchs, L. Hardy and J. Summhammer in [14],
see also papers [15], [16].

\section{General statistical measurement theory}
We start with consideration of general statistical measurement theory. This theory is based 
on the following experimental fact: 

\medskip

(CP) {\it All observed probabilities depend on complexes of experimental physical 
conditions, contexts.}

This principle of {\it contextuality of probabilities} is very general. It is also more or less 
evident (at least after it has been formulated).\footnote{The notion of context is closely related to
so called preparation procedure, see e.g. [8], [17]. However, we prefer context, since context
¨need not be prepared by somebody. A context (e.g., the physical space) can be prepared by Nature.}
It seems that it would be impossible to derive some concrete theoretical results starting with only this
principle. But it is not so. I was surprised by myself that many important consequences can be derived by 
using only this principle, (CP), see [4], [5], cf. Ballentine [6]-[8] and Gudder [9]-[11].
In this paper we would not like to go deeply in technical derivations,
see [4], [5] for the details. We just formulate consequences of the principle (CP). 

We are interested in transformations of probabilities for physical observables due 
to the transition from one context to another. In general statistical measurement theory
a context plays the role of a system of coordinates (statistical coordinates). So we are looking for
transformations of statistical coordinates.

To escape technical complications, 
we consider two physical observables A and B that are dichotomic: 
$A=a_1, a_2$ and $B=b_1, b_2,$ see [4], p. 9967 for the general case of discrete multi 
valued observables. We consider some complex of physical conditions, context, $\cal S$
(`a system of coordinates').
By using $\cal S$ we can prepare a large ensemble $S$ of physical systems. By performing the 
$A$-measurement over elements of $S,$ we get probabilities 
(a kind of vector for `context-coordinate system'): 
$$
p_j^a={\bf P}_{\cal S}(A=a_j), j=1,2.
$$
We now consider contexts ${\cal S}_i^b, i=1,2,$ that realize 
filtrations with respect to the values $B=b_i$ of the $B$-observable. 
Let $S_i^b, i=1,2, $ be corresponding ensembles of physical systems. By performing 
the $A$-measurement over elements of $S_i^b,$ we get probabilities 
$$
p_{ij}^{a/b}={\bf P}_{{\cal S}_i^b}(A=a_j), i,j=1,2.
$$ 
We would like to predict the probabilities $p_j^a$ for the $A$-measurement for an 
arbitrary context $\cal S$ by knowing the probabilities $p_{ij}^{a/b}$ of the $A$-measurement 
for contexts ${\cal S}_i^b, i=1,2, $ corresponding to fixed values $B=b_i$ of the $B$-observable.

We have found [4], [5] that such a prediction rule can always be represented in the form of the transformation:
\begin{equation}
\label {TR1}
p_j^a=p_1 p_{1j}^{a/b}+p_2 p_{2j}^{a/b}+ 2\sqrt{p_1 p_2 p_{1j}^{a/b} p_{2j}^{a/b}}\lambda_j, j=1,2.
\end {equation}
Here $p_i={\bf P}({\cal S}\to {\cal S}_i^b), i=1,2,$ are filtration probabilities.
They are computed in the following way. By applying $B=b_i$ filtration to the ensemble 
$S$ (prepared on the basis of the complex of physical conditions $\cal S$) we get ensembles $S_i^b.$
Then $$p_i\approx \frac{|S_i^b|}{|S|}, i=1,2.$$
when 
$N \to \infty,$ where $N$ is the number of elements in $S.$ We use the symbol $|O|$ to
denote the number of elements in the set $O$.

In transformation (\ref {TR1}) the coefficients $\lambda_j=\lambda_j({\cal S}\to {\cal S}_i^b, A)$ 
are context transition parameters. They give the measure of perturbation of probabilities for 
the observable $A=a_j$ due to transitions from context $\cal S$ to contexts ${\cal S}_i^b.$

{\bf Remark 1.} (Stochasticity of the matrix of transition probabilities). We note that we always have:
\begin{equation}
\label{ST}
p_{11}+p_{12}=1 \; \;{\rm and} \;\; p_{21}+p_{22}=1,
\end {equation}
since $p_{i1}+p_{i2}={\bf P}_{{\cal S}_i^b}(A=a_1)+{\bf P}_{{\cal S}_i^b}(A=a_2)$ and,
for probabilities corresponding to the fixed complex of physical conditions (in our case ${\cal S}_i^b$), 
we use the standard rule for the addition of probabilities of alternatives (in our case $A=a_1$ or $A=a_2).$ 
A matrix ${\bf P}=(p_{ij}^{a/b})$ satisfying (\ref{ST}) is called stochastic matrix, see, e.g. [18].

In [4], [5] we classified statistical measurement theories with respect to magnitudes of
context transition parameters $\lambda_j({\cal S}\to {\cal S}_i^b; A):$

a) trigonometric theory: $|\lambda_j|\leq 1, j=1,2,$ so these parameters can be represented as $\lambda_j=\cos \theta_j, \theta_j \in [0,2 \pi];$

b) hyperbolic theory: $|\lambda_j|>1, j=1,2,$
so $\lambda_j=\pm \cosh \theta_j, \theta_j \in (-\infty, + \infty);$

c) hyper-trigonometric theories: $|\lambda_1|\leq 1$ and $\vert \lambda_2|>1$ or $|\lambda_1|>1$ and $|\lambda_2|\leq 1, $ so $\lambda_1=\cos \theta_1, \lambda_2=\pm \cosh \theta_2$ or $\lambda_1=\pm \cosh \theta_1, \lambda_2=\cos \theta_2.$

We are interested in trigonometric statistical theories. Here we study context transitions
that produce relatively small perturbations of probabilities: 
\begin{equation}
\label{SM}
|\lambda_j({\cal S}\to {\cal S}_i^b; A)|\leq 1,j=1,2.
\end{equation}
In particular, we can consider a statistical measurement theory that operates with classes 
of contexts and observables such that all coefficients $\lambda_j=0.$ Such statistical theories 
of measurement we call {\it classical statistical theories.} In classical theories general transformation, 
(\ref {TR1}), of probabilities corresponding to context transitions is reduced to the well known 
(especially in statistics, see, e.g. [19]) formula of total probability 
\begin{equation}
\label{TP}
p_j^a=p_1 p_{1j}^{a/b}+p_2 p_{2j}^{a/b}
\end {equation}
This formula can easily be derived by using Bayes' formula for conditional probabilities. 
In classical statistical theories we need not pay attention to context dependence of probabilities.
In the mathematical theory we can use one fixed Kolmogorov probability space, [18] (see
the book [21] on extended analysis of possible modifications of the theory of statistical
physical measurements if we use non-Kolmogorovean frameworks to describe statistics of measurements
mathematically - e.g. the framework with context dependent Kolmogorov spaces or von Mises' frequency
framework).
Moreover, we need not 
pay attention under which complex of physical conditions a measurement of $A$ is performed. If you 
like, you can say that $A$ is an objective property of a physical system. Thus classical statistical 
theories are {\it realist} statistical theories.

{\bf Remark 2.} (The formula of total probability in the conventional probabilistic formalism). 
Let $(\Omega, {\cal F}, {\bf P})$ be Kolmogorov's probability space [20], see e.g. [18]. 
Here $\Omega$ is the space of elementary events, $\cal F$ is the $\sigma-$algebra 
of events and $\bf P$- probability measure. Physical observables $A=a_1, a_2$ and $B=b_1, b_2$
are represented by random variables $A(\omega), B(\omega), \omega \in \Omega.$ Conditional probability
is defined by Bayes' formula:
\[{\bf P}(A=a_j/B=b_i)={\frac{{\bf P}(A=a_j, B=b_i)}{{\bf P}(B=b_i)}}.\]
By using additivity of probability we get the formula of total probability:
\begin{equation}
\label{BF}
{\bf P}(A=a_j)={\bf P}(B=b_1) {\bf P}(A=a_j/B=b_1)+
{\bf P}(B=b_2) {\bf P}(A=a_j/B=b_2)
\end{equation}
Here $p_j^a={\bf P}(A=a_j), p_i={\bf P}(B=b_i), p_{ij}^{a/b}={\bf P}(A=a_j/B=b_i).$ 
Here the complex of physical conditions ${\cal S}_i^b$ can be mathematically 
represented as an element of the $\sigma$-algebra of events ${\cal F} -
S_i^b=H_i=\{\omega \in \Omega: B(\omega)=b_i\} \in {\cal F}.$

In statistics events ${\cal S}_i^b$ are considered as statistical hypothesis. Formula (\ref{BF}) is 
used for prediction of the probability of the event $E_j=\{\omega \in \Omega: A(\omega)=a_j\}$ if we
know probabilities of this event under statistical hypothesis $H_i - {\bf P}(E_j/H_i):$
\begin{equation}
\label{BF1}
{\bf P}(E_j)= {\bf P}(H_1) {\bf P}(E_j/H_1) + {\bf P}(H_2) {\bf P}(E_j/H_2).
\end{equation}
However, if perturbations of probabilities corresponding to context transitions 
$$
{\cal S}\to {\cal S}_i^b
$$ 
are relatively large, namely, $\lambda_i \not = 0,$ 
we can not use,  the standard Bayes' rule to predict the probability 
${\bf P}_{{\cal S}}(A=a_j)$ of the event $E_j=\{A=a_j\}$ under the context 
$\cal S$ on the basis of probabilities ${\bf P}_{{\cal S}_i^b}(A=a_j)$ of 
the event $E_j$ under contexts ${\cal S}_i^b.$

{\bf Remark 3.} (Contextual Statistics.) 
It seems natural from the general statistical viewpoint to develop an analogue of 
Bayesian statistical analysis, see Remark 2, for statistical hypothesis 
$H_i$ that could not be represented mathematically as elements of one fixed $\sigma-$algebra $\cal F$ of events.
I tried to find such a statistical formalism in literature; unfortunately, I could not...

In particular
in a trigonometric statistical theory, see (\ref{SM}),
the standard prediction rule (\ref {TP}) is modified:
\begin{equation}
\label {TP2}
p_j^a=p_1 p_{1j}^{a/b}+p_2 p_{2j}^{a/b}+2\sqrt{p_1 p_2 p_{1j}^{a/b} p_{2j}^{a/b}}\cos \theta_j.
\end{equation}
Here `phases' $\theta_j=\theta_j({\cal S}\to {\cal S}_i^b; A)$ have a purely probabilistic meaning. 
These are simply convenient reparametrizations (see further considerations) of perturbation probabilistic 
coefficients $\lambda_j.$ The reader has already recognized in (\ref{TP2}) the form of the quantum rule for 
interference of probabilities. So quantum statistical theory is simply one of trigonometric statistical 
theories of measurement.

However, transformation (\ref{TP2}) is too general for conventional quantum theory
(Dirac-von Neumann formalism). If you perform calculations in the Hilbert space of quantum states,  
\footnote{The origin of the Hilbert space formalism will be discussed later.} you will 
see that quantum transformations (\ref{TP2}) are characterized by the very special choice of 
matrices ${\bf P}=(p_{ij}^{a/b})$ of transition probabilities. Matrices produced by 
quantum formalism are so called {\it double stochastic matrices, } see e.g. A. Peres [22] for the details; 
see (\ref{DS}) for the definition.

\section{The principle of statistical balance in nature}
As we have seen, we could not derive quantum theory only by using the principle of contextuality of 
probabilities, (CP). Well, quantum theory of measurement is one of the statistical theories of measurement. 
We have to find additional conditions that specify quantum statistical theory among all possible statistical
theories of measurement. So, we have to find some specific quantum principles - additional to the general
statistical principle, (CP). We already know that quantum statistical theory
belongs to the class of trigonometric theories. These are theories describing 
classes of contexts and physical observables such that transitions from one context to other produce
relatively small perturbations of probabilities, $|\lambda_i|\leq 1.$

We can formulate this as a {\it principle of relatively small perturbations}, (SP). 

However, (SP) is not, in fact, a physical principle. This is the principle on human investigation of nature. 
We would like to consider theories giving as small as possible errors in prediction. In
principle, we can study some measurements over elementary particles described by the hyperbolic or 
hyper-trigonometric statistical theory. It would simply mean that we consider some observables that are
essentially more sensible to context transitions than quantum observables (we shall discuss this question
later in more details).

In fact, the crucial quantum point is {\it double stochasticity of the matrix of transition probabilities,}
see [4], [5].  The fundamental problem of quantum theory is not interference of probabilities, but:

\medskip

{\bf Why do we have for quantum observables and contexts not only stochasticity condition, 
(\ref{ST}), but also double stochasticity condition? }

\medskip
Namely,
\begin{equation}
 \label{DS}
p_{11}+p_{21}=1 \;\; {\rm{and}} \; \; p_{12} + p_{22}=1.
\end{equation}
As we have seen, stochasticity is easily explained on the basis of general probabilistic arguments. 
But double stochasticity?
\begin{equation}
\label{DDS}
{\bf P}_{{\cal S}_1^b}(A=a_1)+{\bf P}_{{\cal S}_2^b}(A=a_1)=1
\end{equation}
\begin{equation}
\label{DDS1}
{\bf P}_{{\cal S}_1^b}(A=a_2)+{\bf P}_{{\cal S}_2^b}(A=a_2)=1
\end{equation}
Why? I think that equations (\ref{DDS}), (\ref{DDS1}) 
should be interpreted as {\bf statistical balance equations.} If $B=b_1$ preparation produces too much $A=a_1$ 
property, then $B=b_2$ preparation should produce just little $A=a_1$ property: if ${\bf P}_{{\cal S}_1^b}(A=a_1)$
is large, then ${\bf P}_{{\cal S}_2^b}(A=a_1)$ must be small. And vice versa.

We underline that we need not consider A(or B) as an objective property, 
the property of an object-physical system. The $A$(or $B)$ is a statistical property, 
probability for its values are assigned to statistical ensembles corresponding to various contexts. 

Equations (\ref{DDS}), (\ref{DDS1}) are simply the mathematical representation of the great law of nature:
{\it The law of statistical balance.}

By this law in the process of `preparations' for e.g. $B=b_1, b_2$ (that occur randomly in nature) 
the balance between $A=a_1$ and $A=a_2$ properties could not be violated. 
If the law of statistical balance be violated e.g. in the favour of $A=a_1,$ then after some period, 
the $A=a_1$ property would strongly dominate over $A=a_2$ property and, finally, $A=a_2$ property would disappear.
Thus $A$ will be not more a physical observable.

It seems (at least for me) that quantum theory is a statistical measurement theory that is based on 
the principle of statistical balance: 

\medskip

(SB) {\it{Creation of physical properties by `preparations' 
that randomly occur in nature does not violate statistical balance of physical properties.}}

\medskip

On one hand, (SB) is the fundamental principle of quantum theory, the special theory of 
statistical measurements. On the other hand, (SB) is the law of nature. This is the 
very special feature of quantum theory, theory that describes the most fundamental 
`preparations' of physical properties. Such `preparations' are performed on the level of elementary particles. 
The violation of the law of statistical balance on such a level for some property $A$ would imply 
that such a property should sooner or later disappear on micro-level and, as a consequence, on macro-level. 
There could be such properties at the initial stage of the evolution of the universe. However, 
they should disappear. Therefore at the present time we observe only physical properties of 
elementary particles that follow to the law of statistical balance.

Of course, the law of statistical balance in nature can be violated for some contexts 
and (or) observables that are not of the fundamental character. Moreover, we can find observables
that would follow to non-trigonometric transformation laws under the transition from one context
to another. In particular, see [23], hyperbolic rule appears naturally for some generalized observables
represented by POV-measures.

\medskip

{\bf {Conclusion:}} {\it Quantum theory is a statistical theory of measurement 
\footnote{with relatively small perturbations of probabilities corresponding to context transitions} 
based on the principle of statistical balance in nature.}

\medskip

Of course, we do not forget about Planck's constant $h$. This is the constant of energy that could 
not be changed by varying the system of `statistical coordinates' - context (=complex of physical conditions).
This is an analogue of the constant velocity of light in Einstein's theory.

We formulated quantum theory in the way similar to Einstein's formulation of general relativity:

Quantum theory of measurement is a theory about systems of statistical coordinates, contexts, 
and physical observables that are measured in such systems. Numerically these statistical coordinates 
for observables are given by probabilities $p_j^a={\bf P}_{\cal S}(A=a_j).$

This theory gives the possibility to change statistical 
coordinates by using transformation (\ref{TP2}). The trigonometric form of this transition 
law is a consequence of the fact that quantum theory concerns relatively small perturbations of 
statistical coordinates corresponding to transitions from one system of coordinates to other system. 

The fundamental principle of quantum theory is the principle of 
statistical balance in nature. This principle implies that transition-matrices corresponding to 
transformations of statistical coordinates are double stochastic ones. 

It seems that by using this principle and 
the constancy of the level of discretization of energy, Planck constant - we can derive the whole 
quantum formalism. In papers [4], [5] we demonstrated that the Hilbert space probabilistic calculus
can be derived by starting from trigonometric transformation (\ref{TP}).\footnote{As we have remarked,
(\ref{TP}) can be derived in the purely contextual framework - without to apply to superposition or wave arguments.}
The lifting to linear spaces
is based on the recognition in (\ref{TP}) of the well known theorem from elementary geometry -- the (\ref{TP})
is nothing than well known parallelogram theorem:
$$
c^2= a^2+ b^2 + 2ab \cos\theta
$$
with $c^2=p_j^a, a^2 =p_1 p_{1j}^{a/b},  b^2=p_2 p_{2j}^{a/b}.$
Typically the Hilbert space probabilistic calculus is called quantum probabilistic 
calculus. However, it would be more natural to call it simply contextual probabilistic 
calculus. In general it is not coupled rigidly to statistical experiments with elementary
particles. In principle, it could be discovered in purely mathematical investigations a few hundreds
years ago. It is pity that it was discovered only in connection with the creation of quantum statistical
formalism.\footnote{As it was rightly remarked by D. Bohm, it was not so natural at all
to call this formalism -- `mechanics.'} The main quantum discovery -- first experimental and then
theoretical was the discovery of the law of statistical balance in nature or more precisely
-- understanding that there were found fundamental physical observables that should follow to this law.
Unfortunately, this fundamental fact (as we have seen based on the straightforward probabilistic
contextual analysis) was not recognized at the stage of creating of quantum formalism.

{\bf Remark 4.} I would like to underline again that our aim was not to derive the mathematical
formalism of quantum theory. I tried to find fundamental principles of quantum theory. 
For me quantum statistical physics is concentrated in principles (CP), (SP) and (SB).
Manipulations with vectors of Hilbert space is merely mathematics - powerful machinery for calculations
of probabilities.

In this paper we formulated the ``fundamental principle'' of quantum theory by using {\it dichotomous 
observables.} Of coures, if observables take more than two values we could not reduce the distinguishing 
features of quantum probability to {\bf double stochasticity} of matrices of transition probabilities. 
However, it seems that in formulating of a ``fundamental principle'' we could restrict ourselves by considering 
only dichotomous observables -- {\it questions} in terminology of G. Mackey, see [24] for detail;
see also an extended literature on foundations of quantum mechanics from the viewpoint of quantum logic,
e.g., [25].

I would like to remark that I started to be interested in finding of the fundamental quantum
principle after my visit to Atominstitut, Wien, 1997, and intensive discussions with H. Rauch
and J. Summhammer. The important role was played by discussions with A. Zeilinger during the
conference in Venice, 2001, and with C. Fuchs during my visit to Bell' Lab., 2001, and
I. Volovich during my visit to Steklov Math. Institute, 2002, as well as Email exchanges 
with L. Ballentine,  J. Bub, S. Gudder, L. Hardy,  A. Plotnitsky, W. De Baere, W. De Muynck.

\medskip

{\bf References.}

1. A. Zeilinger, On the interpretation and philosophical foundations of
quantum mechanics.
in {\it  Vastakohtien todellisuus.} Festschrift for K.V. Laurikainen.
U. Ketvel et al. (eds), Helsinki Univ. Press, 1996.

2. A. Einstein, {\it Relativity. The special and general theory.} 
New York, Henry Holt, 1920.

3. W. De Muynck, W. De Baere, H. Marten, {\it Found. of Physics,} {\bf 24}, 1589--1663 (1994).

4.  A. Yu. Khrennikov, {\it Linear representations of probabilistic
transformations induced by context transitions.} {\it J. Phys.A: Math. Gen.,} 
{\bf 34}, 9965-9981 (2001); quant-ph/0105059 

A. Yu. Khrennikov, {\it Ensemble fluctuations and the origin of
quantum probabilistic rule.} Reports MSI, V\"axj\"o University, 
{\bf 90}, October (2000).

5. A. Yu. Khrennikov, Ensemble fluctuations and the origin of quantum probabilistic rule.
{\it J. Math. Phys.}, {\bf 43}, N. 2, 789-802 (2002).

A. Yu. Khrennikov, {\it Contextual viewpoint to quantum stochastics.} 
{\it Proc. of Conf. "Quantum Theory: Reconsideration
of Foundations,} series Math. Modeling, {\bf 2 }, V\"axj\"o Univ. Press, 2002;hep-th/0112076. 

[6]  L. E. Ballentine,  Probability theory in quantum mechanics. {\it American
J. of Physics}, {\bf 54,} 883-888 (1986).

[7] L. E. Ballentine,  Interpretations of probability and quantum theory.
Proc. Conf. {\it Foundations of Probability and Physics.}
{\it Q. Prob. White Noise Anal.}, {\bf 13}, 71-84, WSP, Singapore (2001).

[8] L. E. Ballentine, {\it Quantum mechanics.} Englewood Cliffs,
New Jersey, 1989.

[9] S. P. Gudder, Special methods for a generalized probability theory.
{\it Trans. AMS,} {\bf 119}, 428-442 (1965).

[10] S. P. Gudder, {\it Axiomatic quantum mechanics and generalized probability theory.}
Academic Press, New York (1970).

[11] S. P. Gudder, An approach to quantum probability. Proc. Conf.
{\it Foundations of Probability and Physics,} ed. A. Khrennikov.
Quantum Prob. White Noise Anal., {\bf 13}, 147-160, WSP, Singapore (2001).

[12] E. Prugovecki, {\it J. Math. Phys.,} {\bf 7}, 1054-1069 (1966).

[13] E. Prugovecki, {\it Canadian J. Phys.,} {\bf 45}, 2173-2219 (1967).

[14] {\it Proc. of Conf. "Quantum Theory: Reconsideration
of Foundations".} Ed. A. Khrennikov,  series Math. Modeling, {\bf 2 }, V\"axj\"o Univ. Press, 2002;

[15] J. Summhammer, {\em Int. J. Theor. Phys.,} {\bf 33,}  171 (1994) .

[16] L. Hardy (2001), {\it Quantum theory from five reasonable
axioms.} Preprint quant-ph/0101012.

[17]  P. Busch, M. Grabowski, P. Lahti, {\it Operational Quantum Physics.}
Springer Verlag, 1995.

[18]  A. N. Shiryayev, {\it Probability.} Springer, Heidelberg, 1991.
 
[19] H. Cramer, {\it Mathematical theory of statistics.} Univ. Press, Princeton, 1949.

[20]  A. N. Kolmogoroff, {\it Grundbegriffe der Wahrscheinlichkeitsrechnung.}
 (Springer Verlag, Berlin, 1933); reprinted:
 {\it Foundations of the Probability Theory}. 
 (Chelsea Publ. Comp., New York, 1956).

[21]  A.Yu. Khrennikov, {\it Interpretations of Probability.}
VSP Int. Sc. Publishers, Utrecht/Tokyo, 1999.

[22] A.  Peres, {\em Quantum Theory: Concepts and Methods} (Kluwer Academic Publishers) 1994.

[23] A. Yu. Khrennikov, E. Loubents, On relations bewteen probabilities under quantum and classical
measurements. Reports MSI, V\"axj\"o University, March 2002; quant-ph/0204001.

[24] G. W. Mackey, {\it Mathematical foundations of quantum mechanics.}
W. A. Benjamin INc, New York (1963).

[25]  E. Beltrametti  and G. Cassinelli, {\it The logic of Quantum mechanics.}
(Addison-Wesley, Reading, Mass., 1981).

[26] A. Khrennikov, {\it V\"axj\"o interpretation of quantum mechanics.}
Preprint quant-ph/0202107.

\end{document}